# Record-High Superconductivity in Niobium-Titanium Alloy

Jing Guo[1], Gongchang Lin[1,6], Shu Cai[1,6], Chuanying Xi[3], Changjin Zhang[3], Wanshuo Sun[4], Qiuliang Wang[4], Ke Yang[5], Aiguo Li[5], Qi Wu[1], Yuheng Zhang[3], Tao Xiang[1,6]*, Robert J Cava[2]*, and Liling Sun[1,6,7]*

[1]*Institute of Physics, National Laboratory for Condensed Matter Physics, Chinese Academy of Sciences, Beijing 100190, China*
[2]*Department of Chemistry, Princeton University, Princeton, New Jersey 08544, USA*
[3]*High Magnetic Field Laboratory, Chinese Academy of Sciences, Hefei, Anhui 230031, China*
[4]*Institute of Electrical Engineering, Chinese Academy of Sciences, Beijing 100190, China*
[5]*Shanghai Synchrotron Radiation Facilities, Shanghai Institute of Applied Physics, Chinese Academy of Sciences, Shanghai 201204, China*
[6]*University of Chinese Academy of Sciences, Beijing 100190, China*
[7]*Songshan Lake Materials Laboratory, Dongguan, Guangdong 523808, China*

Here we report the observation of extraordinary superconductivity in a pressurized commercial niobium-titanium alloy. We find that its zero-resistance superconductivity persists from ambient pressure to the pressure as high as 261.7 GPa, a record high pressure up to which a known superconducting state can continuously survives. Remarkably, at such an ultra-high pressure, although the ambient pressure volume is shrunk by 45% without structural phase transition, the superconducting transition temperature ($T_C$) increases to ~19.1 K from ~9.6 K, and the critical magnetic field ($H_{C2}$) at 1.8 K has been enhanced to 19 T from 15.4 T. These results set new records for both of the $T_C$ and the $H_{C2}$ among all the known alloy superconductors composed of only transition metal elements. The remarkable high pressure superconducting properties observed in the NbTi alloy not only expand our knowledge on this important commercial superconductor but also are helpful for a better understanding on the superconducting mechanism.

Niobium-titanium (*i.e.* the NbTi alloy near the ratio of 1:1) is a well-known superconductor with a critical transition temperature ($T_C$) of about 10 K at zero magnetic field.[1] Due to its distinguished combined properties of superior high critical magnetic field and high critical supercurrent density, together with its accessible superconducting transition temperature, its affordability, and its easy workability, the NbTi alloy, among thousands of the known superconductors, has been one of the most important material for fabricating commercial superconducting magnets which are widely used in physics, chemistry, materials science and biology.[2-3] However, the knowledge on the properties of this material under extreme conditions is still lack. Understanding the response of superconductors to extreme conditions such as pressure can uncover the unknown properties of the known materials,[4-13] and consequently expand their applications.[4, 14-17] In this study, the new interest on the high pressure studies on the NbTi alloy is stimulated by the discovery of the robust superconductivity in the pressurized high entropy alloy (TaNb)$_{0.67}$(HfZrTi)$_{0.33}$,[14,18] a new class of superconductors which contain elemental niobium, titanium and other transition metal elements.[19,20] How the structural and superconducting properties of the commercial NbTi superconductor responds to the pressure as high as that we applied on the high entropy alloy (HEA) superconductor is of great interest. Here, we experimentally demonstrate that the NbTi alloy has an extraordinary robustness of the superconductivity against pressure up to 261 GPa, a pressure that falls within that of the outer core of the earth, and holds the record-high $T_C$ for the alloys composed of only transition metal elements.

We conducted the high pressure measurements in a non-magnetic diamond anvil cell (DAC) made of BeCu alloy.[11] The samples were obtained from a commercial superconducting NbTi alloy, manufactured by Western Superconducting Technologies Co. Ltd, China. The commercial alloy was composed of a large number of individual NbTi wires surrounded by a CuNi matrix. To obtain a single NbTi superconducting wire, a nitric acid solution was used to remove the wire from the matrix. Before the high pressure measurements, the superconductivity and composition of the sample were characterized. The $T_C$ was about 9.6 K at ambient pressure and its composition was $Nb_{0.44}Ti_{0.56}$. The measured sample was placed in the DAC. Figure 1a shows the temperature dependence of the electrical resistance measured in the pressure range of 1.9- 89.1 GPa for one of the samples. As can be seen, the superconducting transitions of the sample subjected to different pressures are sharp and zero resistance remains present throughout the whole applied pressure range. Looking in detail at the resistance in the low temperature region (Figure 1b), we find that $T_C$ shifts to higher temperature upon increasing pressure. Reproducible results are observed from another sample in the pressure range of 2.9 - 102.7 GPa (see Supplementary Information). A pressure-induced enhancement of $T_C$ below a pressure of 20 GPa was found earlier in a $Nb_{0.53}Ti_{0.47}$ alloy,[21] in good agreement with our low pressure measurements. With the aim of investigating the superconducting behavior at higher pressures, we performed experiments by using diamond anvils with smaller culets (see Supplementary Information) on the third sample over of the wide pressure range of 0.3-261.7 GPa (Figure 1c and 1d). We find, surprisingly, that the superconducting state still survives at a pressure as high as 261.7 GPa, a record-high pressure reported on characterizing superconductivity by transport measurements. Significantly, the zero resistance superconducting state still exists at this high pressure. Looking in detail at the low temperature resistance near the superconducting transition of this sample, we find that its superconducting transition becomes broadened at pressures above 79 GPa (Figure 1d). This broadened transition behavior may be attributed to the inhomogeneous pressure environment that commonly occurs in very high pressure studies.[22] The reproducibility of this unusually robust superconductivity in the NbTi alloy was proved by our high pressure measurements on the fourth sample, subjected to a maximum pressure of 211 GPa before the diamond anvils are shattered (see Supplementary Information).

To investigate the response of superconductivity to magnetic field, we conducted resistance measurements for the pressurized NbTi sample under different magnetic fields at the High Magnetic Field Laboratory in Hefei, China (see Supplementary Information). The pressure-magnetic field-$T_C$ phase diagram for the commercial NbTi alloy, constructed based on the data we obtained from different runs, is plotted in Figure 2. Here we define $T_C$ conservatively as the intersection of the tangent through the inflection point of the resistive transition with a straight-line fit of the normal state resistance just above the transition. At zero magnetic field, $T_C$ displays an increase with pressure and reaches 19 K at ~ 120 GPa, about two times of that seen at ambient pressure. On further compression to 261.7 GPa, a pressure that falls within that of the outer core of the earth,[23] $T_C$ stays almost constant. As shown in the $T_C$ - $B(T)$ panel in

Figure 2, $T_C$ of the pressurized samples shifts to lower temperature upon increasing field over the wide range of pressure. We find that the critical field at 1.8 K for our NbTi sample increases for pressures up to 88.8 GPa and then saturates at higher pressures (Figure2 and Supplementary Information).

To investigate the stability of the crystal structure in the pressurized NbTi alloy, we performed high-pressure X-ray diffraction measurements on the samples at beamline 15U of the Shang Synchrotron Radiation Facility. The results of two independent experiments show that the superconducting NbTi alloy does not undergo a structural phase transition up to the pressure of ~200.5 GPa, maintaining its body-centered-cubic (bcc) structure over the entire pressure range (Figure 3 and Supplementary Information). However, the molar volume decreases dramatically in this wide pressure range: it is compressed by ~43.4% at 200.5 GPa and ~45% (extrapolated from the pressure dependence of unit cell volume) at 261 GPa where the $T_C$ has increased by nearly two times over its ambient pressure value (see panel of -$\Delta V/V_0$ versus pressure of Figure 2).

Figure 4 shows $T_C$ as a function of relative volume of the NbTi superconductor. $T_C$ increases dramatically upon initial compression of the lattice, but saturates when the volume is compressed by more than 34.7 %, corresponding to ~ 120 GPa (Supplementary Information). A similar phenomenon has been observed in the pressurized HEA superconductor,[14] as shown in Fig. 4, while the prominent difference is that the critical pressure $P_C$ for the saturation of $T_C$ is smaller for the HEA. From our study, it is interesting to know that the unusual robustness of the $T_C$ to the dramatic lattice shrinkage found in the NbTi and the HEA superconductors is entirely different from what have been seen in the copper oxide and iron pnictide superconductors, the $T_C$ values of which are very sensitive to the pressure-dependent geometries of the $CuO_2$ planes for the copper oxide superconductors, [24] and the Fe-As-Fe angle or anion height for the iron pnictide superconductors. [25,26]

In addition, the volume dependence of the $T_C$ found in the bcc superconducting NbTi and the HEA superconductors are also quite distinct from that of the bcc superconductor elemental Nb, which is the major elemental constituent of both the NbTi and HEA materials,[27,28] indicating that the high pressure behaviors observed in the NbTi and HEA superconductors do not just emerge from the dilution of Nb. It is of great interest to know whether the superconductivity of the NbTi and HEA materials can withstand pressures higher than the 261 GPa record reported here, but such experiments will be full of challenges.

In conclusion, our results demonstrate that the superconductivity of the commercial NbTi alloy is the most robust among the known superconductors under pressure. Moreover, the values of its $T_C$ and $H_{C2}$ under high pressure set a new record among the alloy superconductors composed of only transition metal elements. Our findings not only reveal the extraordinary high-pressure superconducting properties of the commercial NbTi alloy, but also provide fresh experimental data for a better understanding on the superconducting mechanism.

**Experimental section**

High pressure was generated by a diamond anvil cell made of BeCu alloy with two opposing anvils. A four-probe method was applied for our resistance and magnetoresistance measurements. Diamond anvils with 300 μm, 150 μm and 40 μm culets (flat area of the diamond anvil) were used for different independent measurements, in which the smallest culets of the anvils were employed for the experiments performed at pressures higher than 2 Mbars (1Mbar=100 GPa). Ambient pressure and high pressure magnetoresistance measurements at high magnetic field were performed using the dc-resistive magnet (~35 T) at the High Magnetic Field Laboratory (CHMFL) in Hefei, China. Pressure below and above 60 GPa were determined by the ruby fluorescence method[29] and the pressure dependence of the diamond Raman shift method, respectively.[30,31]

High pressure X-ray diffraction (XRD) measurements were carried out at beamline 15U at the Shanghai Synchrotron Radiation Facility. Low birefringence diamonds with culets of 200 μm and 80 μm were used for the experiments up to ~ 80 GPa and ~ 200.5 GPa, respectively. A monochromatic X-ray beam with a wavelength of 0.6199 Å was employed for all the measurements. The pressures were determined by the ruby fluorescence method[29] and the pressure dependent diamond Raman shift method.[30,31]

**Acknowledgements**

The work in China was supported by the National Key Research and Development Program of China (Grant No. 2017YFA0302900, 2016YFA0300300 and 2017YFA0303103), the NSF of China (Grant Numbers 11427805, U1532267, and 11604376), the Strategic Priority Research Program (B) of the Chinese Academy of Sciences (Grant No. XDB25000000). The work at Princeton was supported by the Gordon and Betty Moore Foundation EPiQS initiative, Grant GBMF-4412. A portion of this work was performed on the Steady High Magnetic Field Facilities, High Magnetic Field Laboratory, CAS.

Correspondence and requests for materials should be addressed to L.S. (llsun@iphy.ac.cn), R.J.C. (rcava@princeton.edu) and T.X. (txiang@iphy.ac.cn)

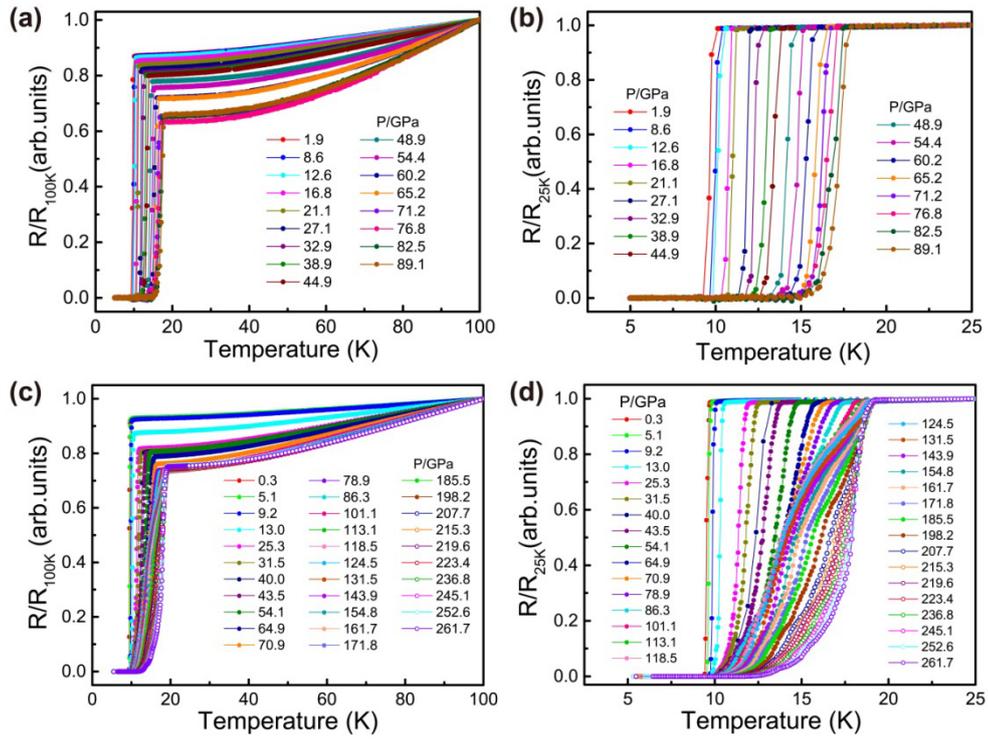

**Figure 1 The superconducting behavior of $Nb_{0.44}Ti_{0.56}$ at high pressures.** (a) Temperature dependence of the resistance in the pressure range of 1.9 GPa to 89.1 GPa. (b) Resistance at lower temperature, exhibiting sharp superconducting transitions with zero resistance and the continuous increase in $T_C$ upon compression. (c) Resistance as a function of temperature for pressures ranging from 0.3 GPa to 261.7 GPa. (d) Resistance versus temperature near the superconducting transition, displaying extraordinary robust superconductivity under very high pressure.

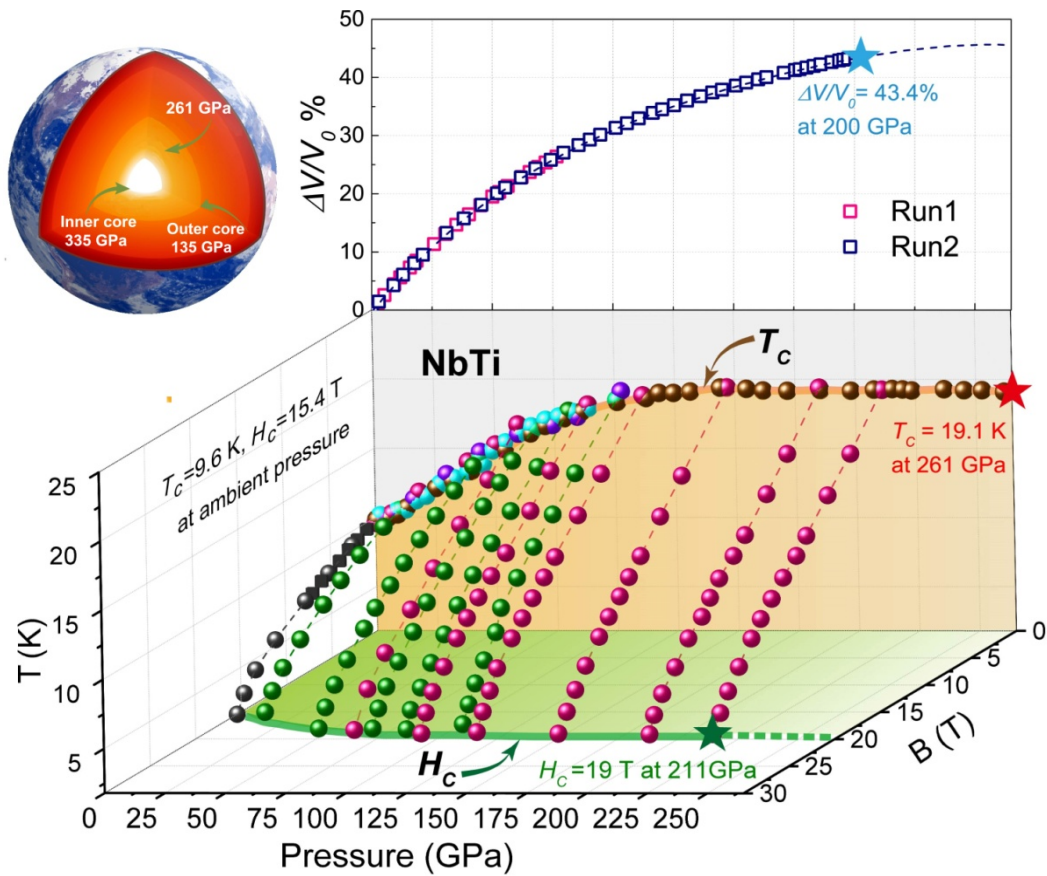

**Figure 2 Superconductivity of $Nb_{0.44}Ti_{0.56}$ under various pressure and magnetic field conditions, and the pressure dependence of its molar volume**. In the panel of pressure versus superconducting transition temperature ($T_C$), the colored balls represent the $T_C$ obtained from the different experimental runs. In the panel of magnetic field, $B(T)$ versus $T_C$, the black, green and red balls represent $T_C$ obtained under zero and applied magnetic fields. In the panel of pressure versus volume ($-\Delta V = V_p - V_0$, where $V_p$ is the volume at fixed pressure and $V_0$ is the ambient-pressure volume), the pink and blue squares represent the results obtained from the two independent runs. The red star labels the $T_C$ value at the record high pressure, the green star marks the critical field at 1.8 K and the maximum pressure of this study and the blue star refers to the relative volume at the highest pressure investigated. The top

left panel displays that the maximum pressure of this study falls in that of outer core of the earth (the image of the earth structure is taken from Ref. [23]).

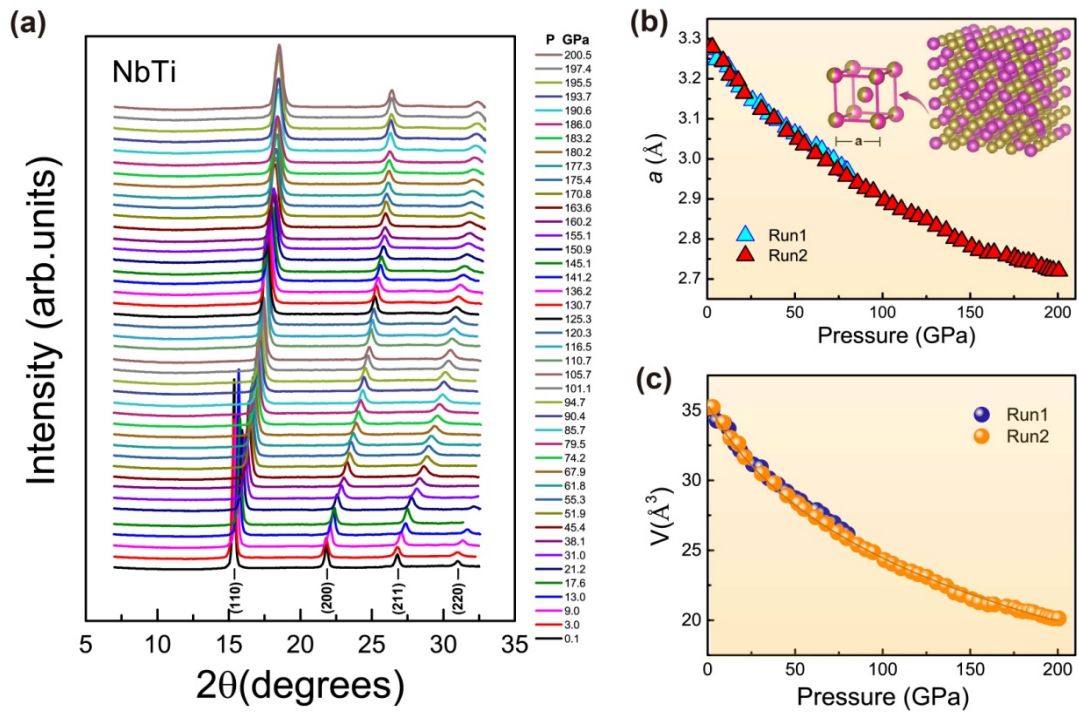

**Figure3 Structure information for NbTi at high pressure.** (a) X-ray powder diffraction patterns collected in the pressure range of 0.1-200.5 GPa. (b) and (c) Pressure dependence of the lattice parameter and unit cell volume for independent two runs. The inset of the figure b displays the schematic crystal structure of the NbTi superconductor.

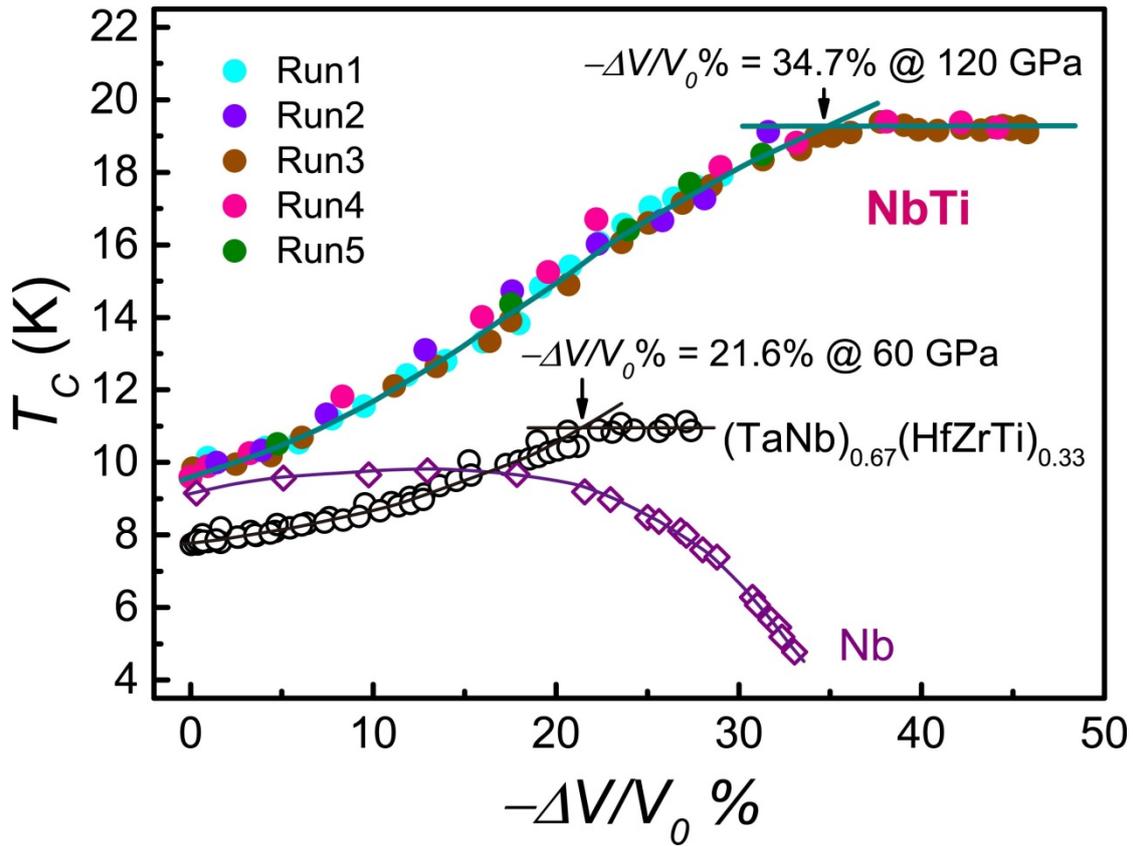

**Figure 4 Plots of the superconducting transition temperature ($T_C$) vs. the change in volume for $Nb_{0.44}Ti_{0.56}$, the high entropy alloy $(TaNb)_{0.67}(HfZrTi)_{0.33}$ and elemental Nb.** The colored solid circles are the data for the NbTi alloy and the open circles are the data for the high entropy alloy, both of them displaying similar high pressure behavior, although with substantially different magnitudes.